\documentclass[onecolumn,showpacs,preprintnumbers,amsmath,amssymb]{revtex4}

\usepackage{epsf}
\usepackage{graphicx}  
\usepackage{dcolumn}   
\usepackage{bm}        
\usepackage{subfigure}

\def\etal{et al.~}
\def\apj{Astrophys. J~}
\def\pasp{Publ. Astron. Soc. Pac.~}
\def\plb{Phys. Lett. B~}

\begin{document}

\title{Galaxy cluster angular size data constraints on dark
energy}

\author{Yun Chen}
 \email{chenyun@mail.bnu.edu.cn}
\affiliation{Department of Astronomy, Beijing Normal University,
Beijing 100875, China}
 \affiliation{Department of Physics, Kansas State University, 116
Cardwell Hall, Manhattan, KS 66506, USA}

\author{Bharat Ratra}
\email{ratra@phys.ksu.edu}
 \affiliation{Department of Physics, Kansas State University, 116
Cardwell Hall, Manhattan, KS 66506, USA}

\date{ \today}

\begin{abstract}

We use angular size versus redshift data for galaxy clusters from
Bonamente \etal \cite{Bonamente2006} to place constraints on model
parameters of constant and time-evolving dark energy cosmological
models. These constraints are compatible with those from other
recent data, but are not very restrictive. A joint analysis of the
angular size data with more restrictive baryon acoustic oscillation
peak length scale and supernova Type Ia apparent magnitude data
favors a spatially-flat cosmological model currently dominated by a
time-independent cosmological constant but does not exclude
time-varying dark energy.

\end{abstract}



\maketitle


\section{INTRODUCTION}
\label{intro}

A number of lines of observational evidence support a ``standard''
model of cosmology with energy budget dominated by far by dark
energy. Dark energy is most simply characterized as a
negative-pressure substance that powers the observed accelerated
cosmological expansion. It can evolve slowly in time and vary weakly
in space, although current data are consistent with it being
Einstein's cosmological constant. On the other hand, some argue that
the observed accelerated expansion should instead be viewed as an
indication that general relativity does not accurately describe
gravitational physics on large cosmological length scales. For
recent reviews see \cite{frieman08, ratra08, caldwell09, sami09,
Bartelmann2010, cai10, brax09}. In what follows we assume that
general relativity provides an accurate description of gravitation
on cosmological length scales.

There are many dark energy models under discussion. For recent
discussions see \cite{wei11, jamil10, maggiore11, dutta10, shao10,
lepe10, sloth10, liu10, honorez10}, and references therein. Perhaps
the most economical --- and the current ``standard'' model
--- is the $\Lambda$CDM model \cite{peebles84}, where the accelerated
cosmological expansion is powered by Einstein's cosmological
constant, $\Lambda$, a spatially homogeneous fluid with equation of
state parameter $\omega_\Lambda = p_\Lambda/\rho_\Lambda = -1$ (with
$p_\Lambda$ and $\rho_\Lambda$ being the fluid pressure and energy
density). In this model the cosmological energy budget is dominated
by $\rho_\Lambda$, with cold dark matter (CDM) being the second
largest contributor. The $\Lambda$CDM model provides a reasonable
fit to most observational constraints, although the ``standard'' CDM
structure formation model might be in some observational trouble
(see, e.g., \cite{Peebles&Ratra2003, Perivolaropoulos2010}).

The $\Lambda$CDM model has a few apparent puzzles. Prominent among
these is that the needed $\Lambda$ energy density scale is of order
an meV, very small compared to the higher (cutoff) value suggested
by a perhaps naive application of quantum mechanics. Another puzzle
is that we happen to be observing at a time not very different from
when the $\Lambda$ energy density started dominating the
cosmological energy budget (this is the ``coincidence'' puzzle).

If the dark energy density --- that responsible for powering the
accelerated cosmological expansion --- slowly decreased in time
(rather than remaining constant like $\rho_\Lambda$), the energy
densities of dark energy and nonrelativistic matter (CDM and
baryons) would remain comparable for a longer period of time, and so
alleviate the coincidence puzzle. Also, a slowly decreasing dark
energy density, that is based on more fundamental physics at a
higher energy density scale much larger than an meV, would result in
a current dark energy density scale of an meV through gradual
decrease over the long lifetime of the Universe. Thus a slowly
decreasing dark energy density could resolve some of the puzzles of
the $\Lambda$CDM model \cite{Ratra&Peebles1988}.

The XCDM parametrization is often used to describe a slowly
decreasing dark energy density. In this parametrization the dark
energy is modeled as a spatially homogenous ($X$) fluid with an
equation of state parameter $w_X = p_X/\rho_X$, where $w_X < -1/3$
is an arbitrary constant and $p_X$ and $\rho_X$ are the pressure and
energy density of the $X$-fluid. When $w_X = -1$, the XCDM
parametrization reduces to the $\Lambda$CDM model, which is a
complete and consistent model. For any other value of $w_X (< -1/3)$
the XCDM parametrization is incomplete as it cannot describe spatial
inhomogeneities (see, e.g., \cite{ratra91}). For computational
simplicity we study the XCDM parametrization only in the
spatially-flat cosmological case.

If the dark energy density evolves in time, physics demands that it
also be spatially inhomogeneous. The $\phi$CDM model --- in which
dark energy is modeled as a scalar field $\phi$ with a gradually
decreasing (in $\phi$) potential energy density $V(\phi)$ --- is the
simplest complete and consistent model of a slowly decreasing (in
time) dark energy density. Here we focus on an inverse power-law
potential energy density $V(\phi) \propto \phi^{-\alpha}$, where
$\alpha$ is a nonnegative constant \cite{Peebles&Ratra1988,
Ratra&Peebles1988}. When $\alpha = 0$ the $\phi$CDM model reduces to
the corresponding $\Lambda$CDM case. Here we only consider the
spatially-flat $\phi$CDM cosmological model.

It has been known for some time now that a spatially-flat
$\Lambda$CDM model with current energy budget dominated by a
constant $\Lambda$ is largely consistent with most observational
constraints (see, e.g., \cite{Jassal10, wilson06, Davis2007,
allen08}). Supernovae Type Ia (SNeIa) apparent magnitude
measurements (e.g., \cite{Riess1998, Perlmutter1999, shafieloo09,
holsclaw10}), in conjunction with cosmic microwave background (CMB)
anisotropy data (e.g., \cite{ratra99, podariu01b, Spergel2003,
Komatsu2009, Komatsu2011}) combined with low estimates of the
cosmological mass density (e.g., \cite{chen03b}), as well as baryon
acoustic oscillation (BAO) peak length scale estimates (e.g.,
\cite{Percival2007, gaztanaga09, samushia09b, wang10a}) strongly
suggest that we live in a spatially-flat $\Lambda$CDM model with
nonrelativistic matter contributing a little less than 30 \% of the
current cosmological energy budget, with the remaining slightly more
than 70 \% contributed by a cosmological constant. These three sets
of data carry by far the most weight when determining constraints on
models and cosmological parameters.

Future data from space missions will tighten the constraints (see,
e.g., \cite{podariu01a, samushia11, wang10b}). However, at present,
it is of great importance to consider independent constraints that
can be derived from other presently available data sets. While these
data do not yet carry as much statistical weight as the SNeIa, CMB
and BAO data, they potentially can reassure us (if they provide
constraints consistent with those from the better known data), or if
the two sets of constraints are inconsistent this might lead to the
discovery of hidden systematic errors or rule out the cosmological
model under consideration.

Other data that have been used to constrain cosmological parameters
include galaxy cluster gas mass fraction (e.g., \cite{allen08,
Samushia&Ratra2008, ettori09}), gamma-ray burst luminosity distance
(e.g., \cite{schaefer07, liang08, wang08, Samushia&Ratra2010}),
large-scale structure (e.g., \cite{courtin11, baldi10,
basilakos10}), strong gravitational lensing (e.g., \cite{chae02,
chae04, lee07, yashar09}), and lookback time (e.g.,
\cite{capozziello04, simon05, Samushia&Ratra2010, dantas11}) or
Hubble parameter (e.g., \cite{Samushia&Ratra2006},
\cite{samushia07}, \cite{fernandez08}, \cite{yang10}) data. While
the constraints from these data are much less restrictive than those
derived from the SNeIa, CMB and BAO data, both types of data result
in largely compatible constraints that generally support a currently
accelerating cosmological expansion. This gives us confidence that
the broad outlines of the ``standard'' cosmological model are now in
place.

Angular size data have also been used to constrain cosmological
parameters (see, e.g., \cite{gurvits99, guerra00, chen03a,
podariu03, khokhlov2011}). In this paper we use the Bonamente \etal
\cite{Bonamente2006}( hereafter B06) galaxy cluster angular size
versus redshift data to derive cosmological constraints. These
measurements were determined from radio and X-ray observations. They
have previously been used to constrain some cosmological parameters
and to test the distance duality relationship of metric gravity
models (see, e.g., \cite{debernardis06, holanda08, lima10, li11,
liang11, meng11}).

In this paper we use the B06 angular size versus redshift data to
constrain cosmological models not previously considered, and to
constrain other cosmological parameters in models previously
considered. We show that these constraints are compatible with those
derived using other data. We also do a joint analysis of this
angular size data and SNeIa and BAO measurements and show that
including the angular size data in the mix affects the constraints,
although not greatly so as the angular size data do not yet have
sufficient weight.

Our paper is organized as follows. In Sec.\ {\ref{equations}} we
present the basic equations of the three dark energy models we
study. Constraints from the B06 angular diameter distances of galaxy
clusters are derived in Sec.\ {\ref{ADD}}. In Sec.\
{\ref{combination}} the BAO data and the SNeIa measurements are used
to constrain the dark energy models. In Sec.{\ref{Joint}} we
determine joint constraints on the dark energy parameters from
different combinations of data sets. Finally, we summarize our main
conclusions in Sec.\ {\ref{summary}}.

%
%
\section{Basic equations of the dark energy models}
\label{equations}

The Friedmann equation of the $\Lambda$CDM model with spatial
curvature can be written as
\begin{equation}
\label{eq:LCDMFriedmann} E^2(z;\textbf{p}) =
\Omega_{m0}(1+z)^3+\Omega_{\Lambda}+(1-\Omega_{m0}-\Omega_{\Lambda})(1+z)^2,
\end{equation}
where $z$ is the redshift, $E(z) = H(z)/H_0$ is the dimensionless
Hubble parameter where $H_0$ is the Hubble constant, and the
model-parameter set is $\textbf{p} = (\Omega_{m0},
\Omega_{\Lambda})$ where $\Omega_{m0}$ is the nonrelativistic
(baryonic and cold dark) matter density parameter and
$\Omega_{\Lambda}$ that of the cosmological constant. Throughout,
the subscript $0$ denotes the value of a quantity today. In this
paper, the subscripts $\Lambda$, $X$ and $\phi$ represent the
corresponding quantities of the dark energy component in the
$\Lambda$CDM, XCDM and $\phi$CDM models.

In this work, for computational simplicity, the spatial curvature is
set to zero in the XCDM and $\phi$CDM cases. Then the Friedmann
equation for the XCDM parametrization is
\begin{equation}
\label{eq:XCDMFriedmann} E^2(z;\textbf{p}) =
\Omega_{m0}(1+z)^3+(1-\Omega_{m0})(1+z)^{3(1+w_X)},
\end{equation}
where the model-parameter set is $\textbf{p} = (\Omega_{m0}, w_X)$.

In the $\phi$CDM model, the inverse power law potential energy
density of the scalar field adopted in this paper is $V(\phi) =
\kappa m_p^2 \phi^{-\alpha}$, where $m_p$ is the Planck mass, and
$\alpha$ and $\kappa$ are non-negative constants
\cite{Peebles&Ratra1988}. In the spatially-flat case the Friedmann
equation of the $\phi$CDM model is
\begin{equation}
\label{eq:phiCDMFriedmann} H^2(z;\textbf{p}) =
\frac{8\pi}{3m_p^2}(\rho_m + \rho_{\phi}),
\end{equation}
where $H(z) = \dot{a}/a$ is the Hubble parameter, and $a(t)$ is the
cosmological scale factor and an overdot denotes a time derivative.
The energy densities of the matter and the scalar field  are
\begin{equation}
\label{eq:rhom} \rho_m = \frac{m_p^2}{6\pi}a^{-3},
\end{equation}
and
\begin{equation}
\label{eq:rhophi} \rho_{\phi} = \frac{m_p^2}{32\pi} (\dot{\phi}^2 +
\kappa m_p^2 \phi^{-\alpha}),
\end{equation}
respectively. According to  the definition of the dimensionless
density parameter, one has
\begin{equation}
\label{eq:Omegam} \Omega_m(z) = \frac{8\pi \rho_m}{3m_p^2H^2} =
\frac{\rho_m}{\rho_m + \rho_{\phi}}.
\end{equation}
The scalar field $\phi$ obeys the differential equation
\begin{equation}
\label{eq:dotphi} \ddot{\phi} + 3 \frac{\dot{a}}{a}\dot{\phi} -
\frac{\kappa \alpha}{2} m_p^2 \phi^{-(\alpha+1)} = 0.
\end{equation}
Using Eqs.\ (\ref{eq:phiCDMFriedmann}) and (\ref{eq:dotphi}), as
well as the initial conditions described in
\cite{Peebles&Ratra1988}, one can numerically compute the Hubble
parameter $H(z)$. In this case, the model-parameter set is
$\textbf{p} = (\Omega_{m0}, \alpha)$.

To use observational data to constrain cosmological models, we need
various distance expressions. The coordinate distance is
\begin{equation}
\label{eq:r} r = \frac{c}{a_0 H_0 \sqrt{|\Omega_k|}} {\rm
sinn}\left[\sqrt{|\Omega_k|} \int_{0}^{z}\frac{dz'}{E(z')}\right],
\end{equation}
where $\Omega_k$ is the spatial curvature density parameter and $c$
is the speed of light, and
\begin{equation}
\frac{{\rm
sinn}(\sqrt{|\Omega_k|}\;x)}{\sqrt{|\Omega_k|}}=\begin{cases}
{\sin(\sqrt{-\Omega_k}\;x)}/{\sqrt{-\Omega_k}} & (\Omega_k<0),\\
x & (\Omega_k=0), \\
{\sinh(\sqrt{\Omega_k}\;x)}/{\sqrt{\Omega_k}} & (\Omega_k>0).
\end{cases}
\end{equation}
The luminosity distance $d_L$ and the angular diameter distance
$d_A$ are simply related to the coordinate distance as
\begin{equation}
\label{eq:dL} d_L = (a_0 r)(1+z),
\end{equation}
and
\begin{equation}
\label{eq:dA} d_A = (a_0 r)/(1+z).
\end{equation}

\section{Constraints from the angular size data}
\label{ADD}

X-ray observations of the intracluster medium combined with radio
observations of the galaxy cluster Sunyaev-Zeldovich effect allow
for an estimate of the angular diameter distance (ADD) $d_A$ of
galaxy clusters. Here we use the 38 ADDs of B06 to constrain
cosmological parameters. These data can be found in Tables 1 and 2
of B06. For convenience, we re-collect them in Table \ref{tab:ADD}.

\begin{table}
\vspace*{-12pt}
\begin{center}
\begin{tabular}{lcl}
\hline\hline
Cluster & $z$ & $d_A$\;(Mpc)\\
\hline
 Abell 1413
 & 0.142
 & $780^{+180}_{-130}$\\
 Abell  2204
 & 0.152
 & $610^{+60}_{-70}$\\
Abell 2259
 & 0.164
 & $580^{+290}_{-250}$\\
Abell 586
 & 0.171
& $520^{+150}_{-120}$\\
Abell 1914 & 0.171
 & $440^{+40}_{-50}$\\
Abell 2218
 & 0.176
 & $660^{+140}_{-110}$\\
Abell 665 & 0.182
& $660^{+90}_{-100}$\\
Abell 1689 &0.183
& $650^{+90}_{-90}$\\
Abell 2163 &0.202
& $520^{+40}_{-50}$\\
Abell 773 &0.217
& $980^{+170}_{-140}$\\
Abell 2261 &0.224
& $730^{+200}_{-130}$\\
Abell 2111 &0.229
& $640^{+200}_{-170}$\\
Abell 267 &0.230
& $600^{+110}_{-90}$\\
RX J2129.7+0005 &0.235
& $460^{+110}_{-80}$\\
Abell 1835 &0.252
& $1070^{+20}_{-80}$\\
Abell 68 &0.255
& $630^{+160}_{-190}$\\
Abell 697 &0.282
& $880^{+300}_{-230}$\\
Abell 611 &0.288
& $780^{+180}_{-180}$\\
ZW 3146 &0.291
& $830^{+20}_{-20}$\\
Abell 1995 &0.322
& $1190^{+150}_{-140}$\\
MS 1358.4+6245 &0.327
& $1130^{+90}_{-100}$\\
Abell 370 &0.375
& $1080^{+190}_{-200}$\\
MACS J2228.5+2036 &0.412
& $1220^{+240}_{-230}$\\
RX J1347.5-1145 &0.451
& $960^{+60}_{-80}$\\
MACS J2214.9-1359 &0.483
& $1440^{+270}_{-230}$\\
MACS J1311.0-0310 &0.490
& $1380^{+470}_{-370}$\\
CL 0016+1609 &0.541
& $1380^{+220}_{-220}$\\
MACS J1149.5+2223 &0.544
& $800^{+190}_{-160}$\\
MACS J1423.8+2404 &0.545
& $1490^{+60}_{-30}$\\
MS 0451.6-0305
 &0.550
 & $1420^{+260}_{-230}$\\
MACS J2129.4-0741
 &0.570
 & $1330^{+370}_{-280}$\\
MS 2053.7-0449
 &0.583
 & $2480^{+410}_{-440}$\\
 MACS J0647.7+7015
 &0.584
 & $770^{+210}_{-180}$\\
 MACS J0744.8+3927
 &0.686
 & $1680^{+480}_{-380}$\\
MS 1137.5+6625
 &0.784
 & $2850^{+520}_{-630}$\\
RX J1716.4+6708
 &0.813
 & $1040^{+510}_{-430}$\\
MS 1054.5-0321
 &0.826
 & $1330^{+280}_{-260}$\\
CL J1226.9+3332
 &0.890
 & $1080^{+420}_{-280}$\\
\hline
\end{tabular}
\end{center}
\caption{Angular diameter distances of galaxy clusters from
B06.}\label{tab:ADD}
\end{table}

There are three sources of uncertainty in the measurement of $d_A$:
the cluster modeling error $\sigma_{\rm mod}$; the statistical error
$\sigma_{\rm stat}$; and, the systematic error $\sigma_{\rm sys}$.
The modeling errors are shown in Table \ref{tab:ADD} and the
statistical and systematic errors are presented in Table 3 of B06.
In our analysis here we combine these errors in quadrature. Thus,
the total uncertainty $\sigma_{\rm tot}$ satisfies $\sigma_{\rm
tot}^2 = \sigma_{\rm mod}^2 + \sigma_{\rm stat}^2 + \sigma_{\rm
sys}^2$.

We constrain cosmological parameters by minimizing $\chi_{ADD}^2$,
\begin{equation}
\label{eq:chi2ADD} \chi_{ADD}^2 (H_0, \textbf{p}) =
\sum_{i=1}^{38}\frac{[d^{\rm th}_A (z_i; H_0, \textbf{p})-d_A^{\rm
obs}(z_i)]^2}{\sigma^2_{{\rm tot},i}}.
\end{equation}
Here $z_i$ is the redshift of the observed galaxy cluster, $d_A^{\rm
th}$ is the predicted value of the ADD in the cosmological model
under consideration and $d_A^{\rm obs}$ is the measured value. From
$\chi_{ADD}^2 (H_0, \textbf{p})$ we compute the likelihood function
$L(H_0, \textbf{p})$. We then treat $H_0$ as a nuisance parameter
and marginalize over it using a gaussian prior with $H_0 = 68\pm3.5$
km s$^{-1}$ Mpc$^{-1}$ \cite{chen03} to get a likelihood function
$L(\textbf{p})$ that is a function of only the cosmological
parameters of interest. The best-fit parameter values $p*$ are those
that maximize the likelihood function and the 1, 2 and 3 $\sigma$
constraint contours are the set of cosmological parameters (centered
on $p*$) that enclose 68.27, 95.45 and 99.73 \%, respectively, of
the probability under the likelihood function.

\begin{figure}[t]
\centering
  \includegraphics[angle=0,width=90mm]{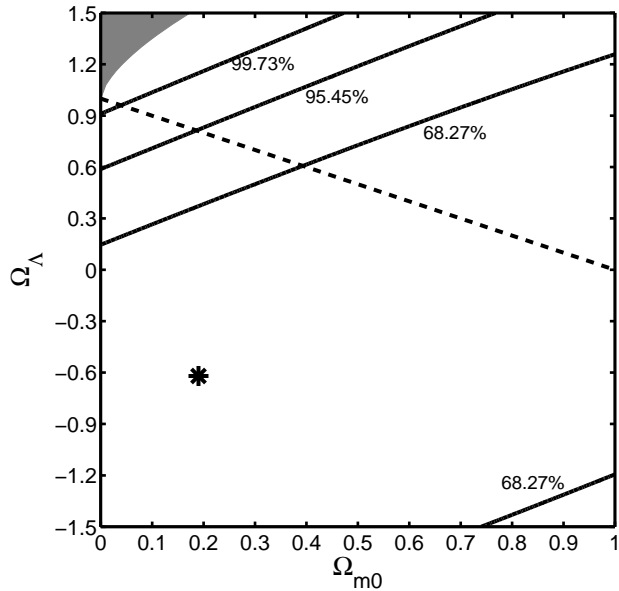}

\caption{
1, 2, and 3 $\sigma$ constraint contours for the $\Lambda$CDM model
from the ADD data. The dashed diagonal line corresponds to
spatially-flat models and the shaded area in the upper left-hand
corner is the region for which there is no big bang. The star marks
the best-fit pair $(\Omega_{m0}, \Omega_{\Lambda}) = (0.19, -0.62)$
with $\chi^2_{\rm min}=30.1$.
} \label{fig:LCDM_ADD}
\end{figure}

\begin{figure}[t]
\centering
  \includegraphics[angle=0,width=90mm]{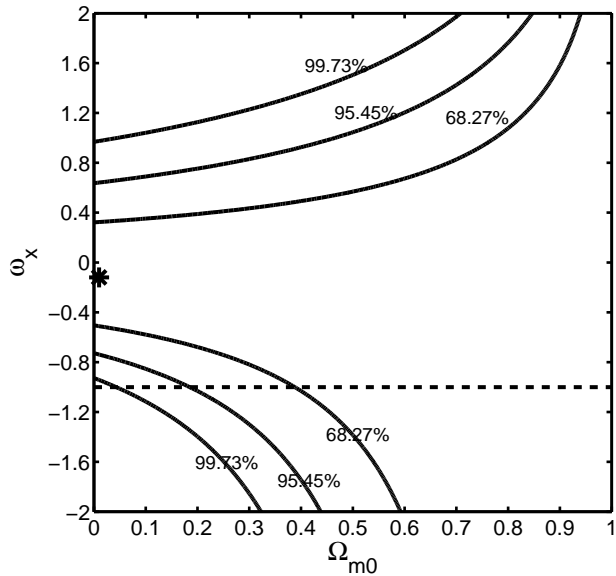}

\caption{
1, 2, and 3 $\sigma$ constraint contours for the XCDM
parametrization from the ADD data. The dashed horizontal line at
$\omega_X = -1$ corresponds to spatially-flat $\Lambda$CDM models.
The star marks the best-fit pair $(\Omega_{m0}, w_X) = (0.01,
-0.12)$ with $\chi^2_{\rm min}=30.2$.
} \label{fig:XCDM_ADD}
\end{figure}

\begin{figure}[t]
\centering
  \includegraphics[angle=0,width=90mm]{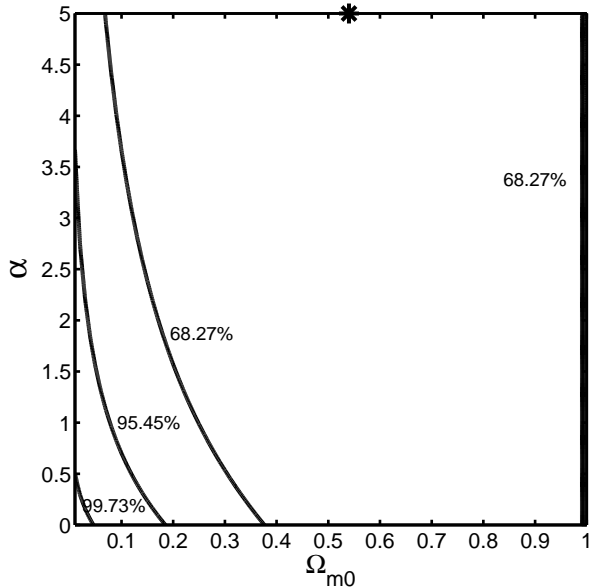}

\caption{
1, 2, and 3 $\sigma$ constraint contours for the $\phi$CDM model
from the ADD data. The horizontal axis at $\alpha = 0$ corresponds
to spatially-flat $\Lambda$CDM models. The star marks the best-fit
pair $(\Omega_{m0}, \alpha) = (0.54, 5)$ with $\chi^2_{\rm
min}=37.3$.
} \label{fig:phiCDM_ADD}
\end{figure}

Figures \ref{fig:LCDM_ADD}---\ref{fig:phiCDM_ADD} show the
constraints from the ADD data on the three dark energy models we
consider. Comparing these results to those shown in Figs.\ 1---3 of
\cite{chen03a}, derived using the compact radio source angular size
data of Gurvits \etal \cite{gurvits99}, and to Figs.\ 1---2 of
\cite{podariu03}, derived using the FRIIb radio galaxy angular size
data from Guerra \etal \cite{guerra00}, we see that the B06 galaxy
cluster angular size data result in approximately comparable
constraints on cosmological parameters as those from the two earlier
angular size data sets. These ADD constraints are comparable to
those from gamma-ray burst data (\cite{Samushia&Ratra2010},
Figs.\ 1---3 and 7---9) as well as those from Hubble parameter
measurements (\cite{samushia07}, Figs.\ 1---3).

Clearly, current ADD data constraints are not very restrictive,
although it is encouraging that the ADD constraints on these dark
energy models do not disfavor the regions of parameter space that
are favored by other data. More importantly, we anticipate that near
future ADD data will provide significantly more restrictive
constraints on cosmological parameters.

\section{Constraints from BAO and SNeIa data}
\label{combination}

The BAO peak length scale can be used as a standard ruler to
constrain cosmological parameters. Here we use the recent Percival
et al.\ \cite{Percival2010} BAO data to constrain parameters of
the $\Lambda$CDM and $\phi$CDM models and the XCDM parametrization.

Percival \etal (2010) measure the position of the BAO peak from the
SDSS DR7 and 2dFGRS data, determining
$r_s(z_d)/D_V(z=0.275)=0.1390\pm 0.0037$, where $r_s(z_d)$ is the
comoving sound horizon at the baryon drag epoch, and $D_V(z)\equiv
[(1+z)^2 d_A^2 cz/H(z)]^{1/3}$. By using $\Omega_{m0}
h^2=0.1326\pm0.0063$ and $\Omega_{b0}h^2=0.02273\pm0.00061$ (here
$\Omega_{b0}$ is the current value of the baryonic mass density
parameter and $h$ is the Hubble constant in units of 100 km s$^{-1}$
Mpc$^{-1}$) from \textit{WMAP5} \cite{Komatsu2009}, one can
further get
\begin{equation}
\label{eq:DV}
D_V(0.275)=(1104\pm30)\left(\frac{\Omega_{b0}h^2}{0.02273}\right)^{-0.134}\left(\frac{\Omega_{m0}
h^2}{0.1326}\right)^{-0.255}\ {\rm Mpc},
\end{equation}
as is shown in Eq.\ (13) of \cite{Percival2010}. The error on
$\Omega_{b0} h^2$ is ignored in this work, as the \textit{WMAP5}
data constrains $\Omega_{b0} h^2$ to 0.5 \%.

\begin{figure}[t]
\centering
  \includegraphics[angle=0,width=90mm]{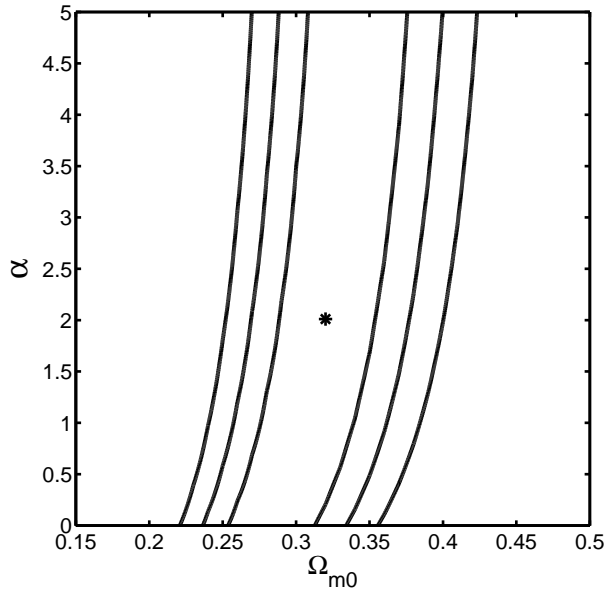}

\caption{
1, 2, and 3 $\sigma$ constraint contours for the $\phi$CDM model
from the BAO data. The horizontal axis at $\alpha = 0$ corresponds
to spatially-flat $\Lambda$CDM models. The star marks the best-fit
pair $(\Omega_{m0}, \alpha) = (0.32, 2.01)$ with $\chi^2_{\rm
min}=0.169$. } \label{fig:phiCDM_BAO}
\end{figure}

Our results for the $\Lambda$CDM model and the XCDM parametrization
agree very well with the Ref.\ \cite{Percival2010} results shown in
their Fig.\ 5. Our results for the $\phi$CDM model are shown in
Fig.\ \ref{fig:phiCDM_BAO}. BAO data primarily constrains
$\Omega_{m0}$ significantly, leaving $\Omega_{\Lambda}$, $w_X$ and
$\alpha$ almost unconstrained (see, e.g., \cite{samushia09a}).
The results from the BAO data \cite{Percival2010} are most
directly comparable to those derived from the earlier BAO data of
Ref.\ \cite{Eisenstein2005}. Comparing to the constraints shown in
Figs.\ 2---4 of Ref.\ \cite{samushia09a}, one sees that the
Percival et al.\ (2010) data results in slightly more restrictive
constraints than the Eisenstein et al.\ (2005) data.

Type Ia supernovae are standardizable candles that can be used to
constrain cosmological parameters. Here we use the recent Union2
compliation of 557 SNeIa (covering a redshift range $0.015\leq z
\leq 1.4$) of Amanullah et al.\ \cite{Amanullah2010} to constrain parameters of
the $\Lambda$CDM and $\phi$CDM models and the XCDM parametrization.

Cosmological constraints from SNeIa data are obtained by using the
distance modulus $\mu (z)$. The theoretical (predicted) distance
modulus is
\begin{equation}
\label{eq:mu} \mu_{\rm th}(z; \textbf{p}, \mu_0 )=5\log_{10} [D_L(z;
\textbf{p})]+\mu_0,
\end{equation}
where $\mu_0=42.38-5\log_{10}h$ and the Hubble-free luminosity
distance is given by
\begin{equation}
\label{eq:DL} D_L(z;\textbf{p})=\frac{H_0}{c}d_L=(1+z)\int_0^z
\frac{dz'}{E(z';\textbf{p})}.
\end{equation}
The best-fit values of cosmological model parameters can be
determined by minimizing the $\chi^2$ function
\begin{equation}
\label{eq:chi2SN}
\chi^2_{SN}(\textbf{p},\mu_0)=\sum_{i=1}^{557}\frac{[\mu_{{\rm
th},i}(z_i; \textbf{p},\mu_0)-\mu_{{\rm
obs},i}(z_i)]}{\sigma^2_{\mu_i}},
\end{equation}
where $\mu_{{\rm obs},i}(z_i)$ is the distance modulus obtained from
observations and $\sigma_{\mu_i}$ is the total uncertainty of the
SNeIa data. The zero-point $\mu_0$ is treated as a nuisance
parameter and marginalized over analytically \cite{DiPietro&Claeskens2003,
Perivolaropoulos2005, Nesseris&Perivolaropoulos2005}. The covariance matrix with or without systematic errors can
be found on the web (http://supernova.lbl.gov/Union).

\begin{figure}[t]
\centering
  \includegraphics[angle=0,width=90mm]{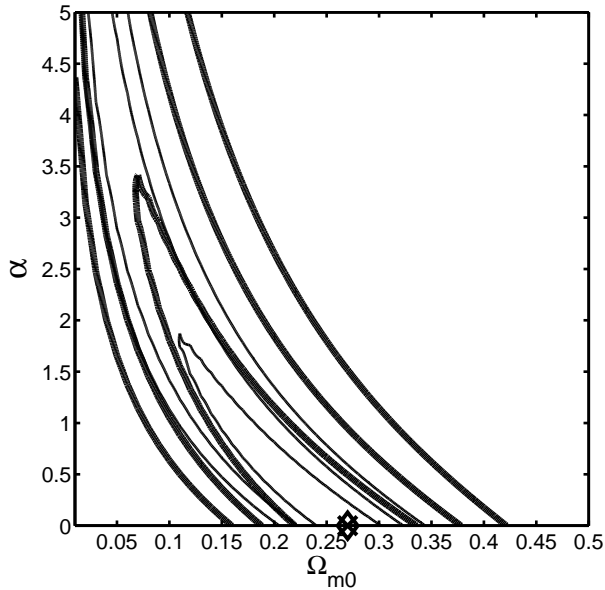}

 \caption{
1, 2, and 3 $\sigma$ constraint contours for the $\phi$CDM model
from the SNeIa data. The horizontal axis at $\alpha = 0$ corresponds
to spatially-flat $\Lambda$CDM models. Thin solid lines (best fit at
$\Omega_{m0} = 0.27$ and $\alpha = 0.0$ with $\chi^2_{\rm min}=543$,
marked by ``$\times$'' ) exclude systematic errors, while thick
solid lines (best fit at $\Omega_{m0} = 0.27$ and $\alpha = 0.0$
with $\chi^2_{\rm min}=531$, marked by ``$\diamondsuit$'') account
for systematics.} \label{fig:phiCDM_SN}
\end{figure}

Our results for the $\Lambda$CDM model and the XCDM parametrization
agree very well with the Ref.\ \cite{Amanullah2010} results shown in
their Figs.\ 10 and 11. The constraints on $\phi$CDM model
parameters from these data are shown in Fig.\ \ref{fig:phiCDM_SN}.
Comparing to Fig.\ 4 of Ref.\ \cite{samushia09b}, we see that the
constraints from the Ref.\ \cite{Amanullah2010} data with systematics
errors are approximately comparable to those from the earlier
Ref.\ \cite{Kowalski2008} data for 307 SNeIa without systematic
errors. Like the case for the BAO data, the SNeIa data constraints
are also fairly one dimensional, tightly constraining one
combination of the cosmological parameters while only weakly
constraining the ``orthogonal'' combination.


\section{Joint constraints}
\label{Joint}

\begin{figure}[t]
\centering
  \includegraphics[angle=0,width=90mm]{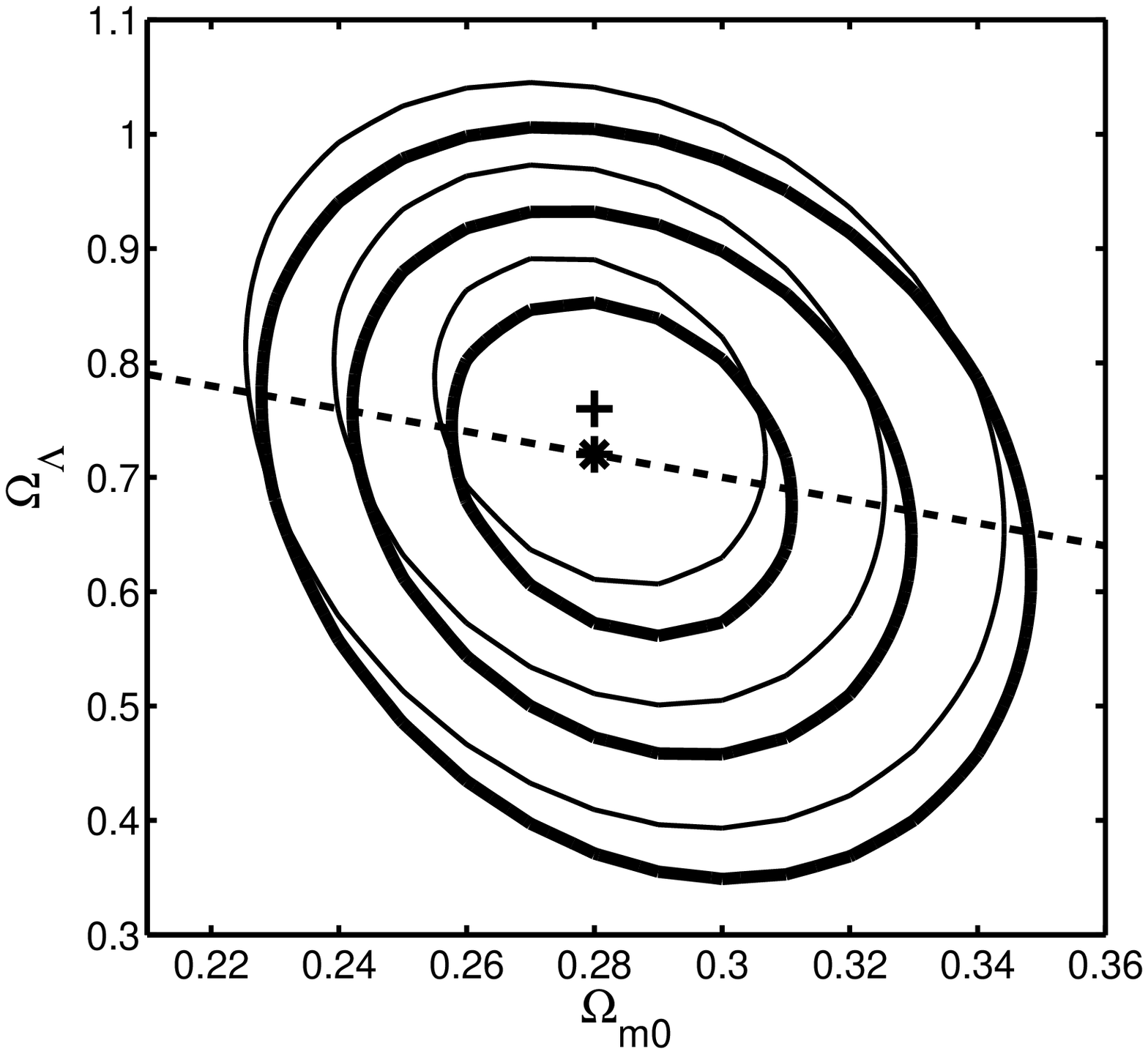}

 \caption{
Thick (thin) solid lines are 1, 2, and 3 $\sigma$ constraint
contours for the $\Lambda$CDM model from a joint analysis of the BAO
and SNeIa (with systematic errors) data, with (and without) the ADD
data. The cross (``+'') marks the best-fit point determined from the
joint sample without the ADD data at $\Omega_{m0}=0.28$ and
$\Omega_{\Lambda}=0.76$ with $\chi^2_{\rm min}=531$. The star
(``$*$'') marks the best-fit point determined from the joint sample
with the ADD data at $\Omega_{m0}=0.28$ and $\Omega_{\Lambda}=0.72$
with $\chi^2_{\rm min}=565$. The dashed sloping line corresponds to
spatially-flat models.} \label{fig:LCDM_com}
\end{figure}

\begin{figure}[t]
\centering
  \includegraphics[angle=0,width=90mm]{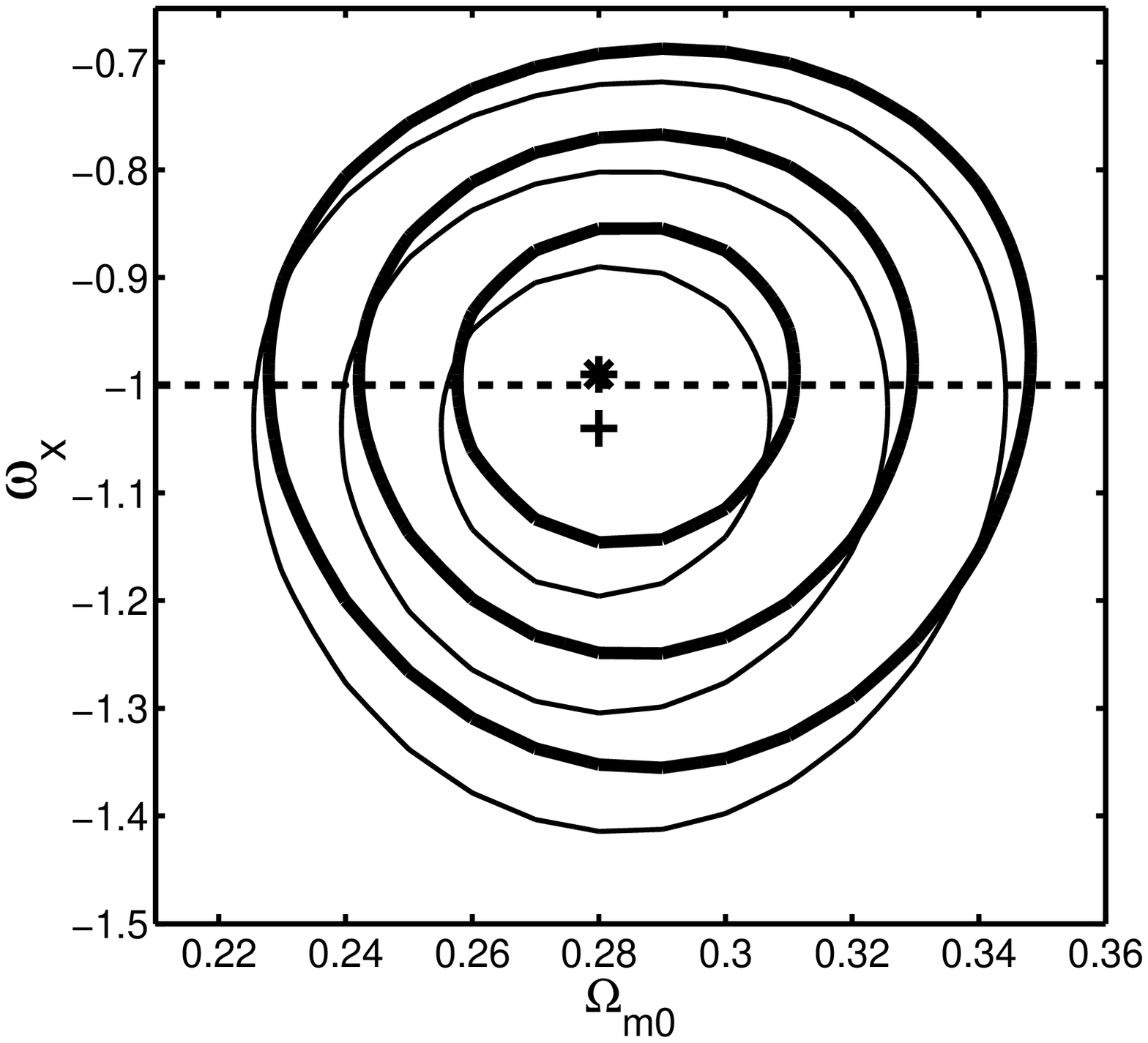}

 \caption{
Thick (thin) solid lines are 1, 2, and 3 $\sigma$ constraint
contours for the XCDM parametrization from a joint analysis of the
BAO and SNeIa (with systematic errors) data, with (and without) the
ADD data. The cross (``+'') marks the best-fit point determined from
the joint sample without the ADD data at $\Omega_{m0}=0.28$ and
$\omega_X=-1.04$ with $\chi^2_{\rm min}=531$. The star (``$*$'')
marks the best-fit point determined from the joint sample with the
ADD data at $\Omega_{m0}=0.28$ and $\omega_X=-0.99$ with
$\chi^2_{\rm min}=565$. The dashed horizontal line at $\omega_X =
-1$ corresponds to spatially-flat $\Lambda$CDM models.}
\label{fig:XCDM_com}
\end{figure}

\begin{figure}[t]
\centering
  \includegraphics[angle=0,width=90mm]{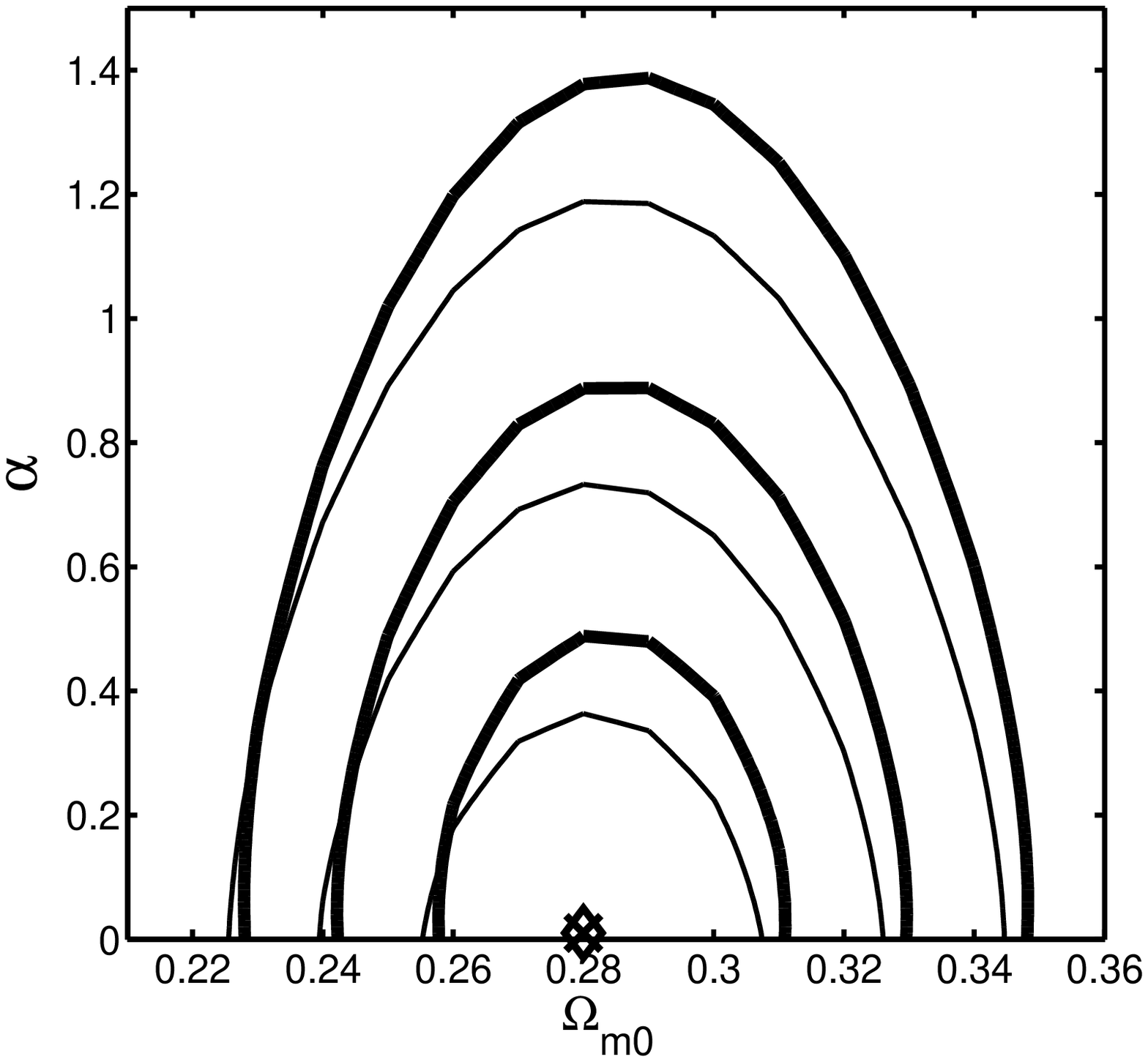}

 \caption{
Thick (thin) solid lines are 1, 2, and 3 $\sigma$ constraint
contours for the $\phi$CDM model from a joint analysis of the BAO
and SNeIa (with systematic errors) data, with (and without) the ADD
data. The cross (``$\times$'') marks the best-fit point determined
from the joint sample without the ADD data at $\Omega_{m0}=0.28$ and
$\alpha=0$ with $\chi^2_{\rm min}=531$. The diamond
(``$\diamondsuit$'') marks the best-fit point determined from the
joint sample with the ADD data at $\Omega_{m0}=0.28$ and
$\alpha=0.01$ with $\chi^2_{\rm min}=572$. The $\alpha = 0$
horizontal axis corresponds to spatially-flat $\Lambda$CDM models.}
\label{fig:phiCDM_com}
\end{figure}

Figures \ref{fig:LCDM_com}---\ref{fig:phiCDM_com} show the
constraints on the cosmological parameters for the $\Lambda$CDM and
$\phi$CDM models and the XCDM parametrization from a joint analysis
of the BAO and SNeIa data, as well as from a joint analysis of the
BAO, SNeIa and ADD data. With the inclusion of systematic errors in
the analysis of the SNeIa data of Ref.\ \cite{Amanullah2010}, the new
joint BAO and SNeIa constraints (thin solid contours in Figs.\
\ref{fig:LCDM_com}---\ref{fig:phiCDM_com}) are similar to the
earlier ones shown in Figs.\ 4---6 of Ref.\ \cite{Samushia&Ratra2010} that
made use of the smaller SNeIa data set of Ref.\ \cite{Kowalski2008}
that did not include systematic errors.

\begin{figure}[t]
\centering
 $\begin{array}{cc}
\includegraphics[width=0.5\textwidth]{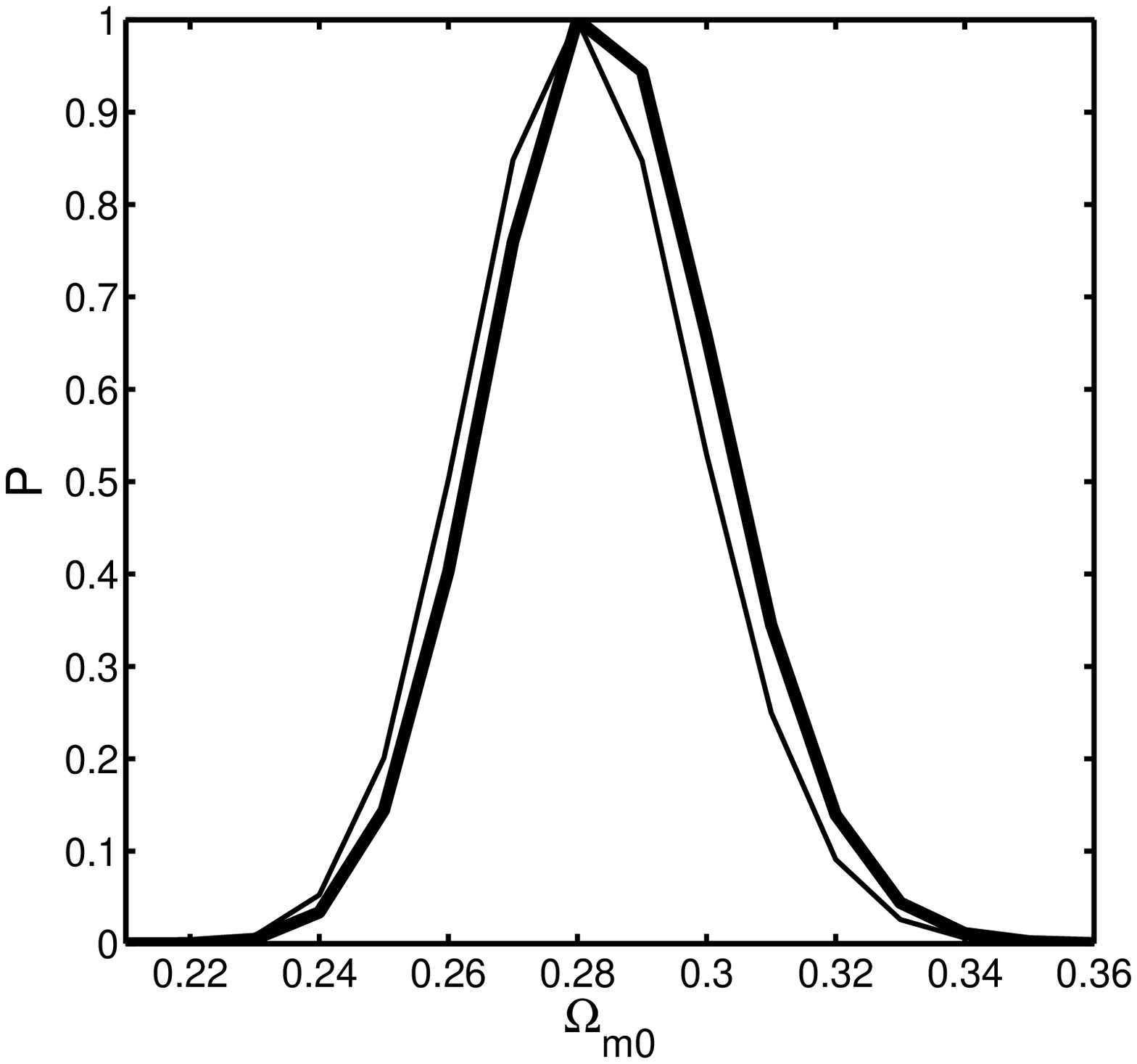}\\
\includegraphics[width=0.5\textwidth]{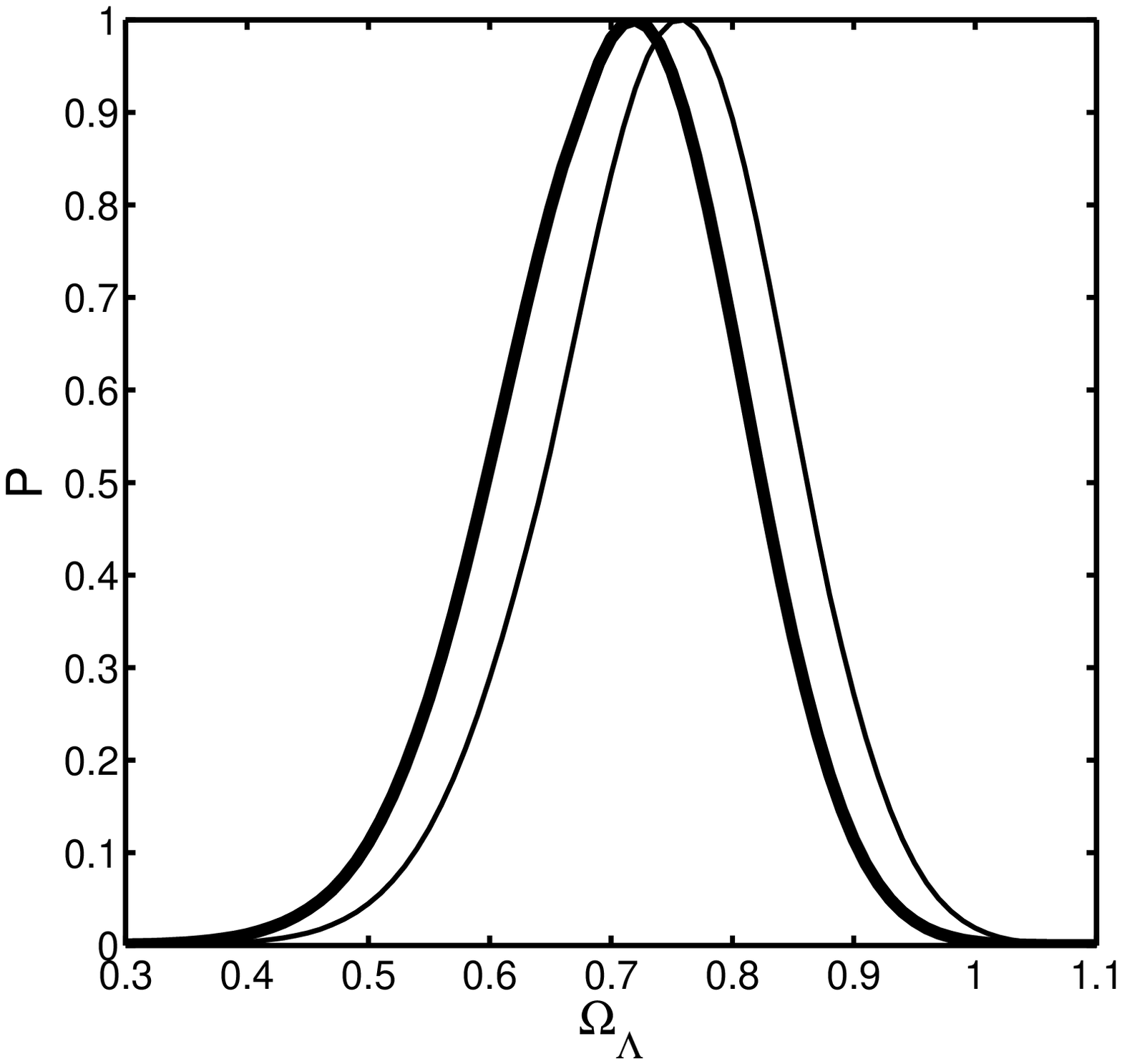}
\end{array}$
 \caption{One-dimensional marginalized distribution probabilities
of the cosmological parameters for the LCDM model. Thick (thin)
lines are results from a joint analysis of the BAO and SNeIa (with
systematic errors) data, with (and without) the ADD data.}
\label{fig:Pro_LCDM}
\end{figure}

\begin{figure}[t]
\centering $\begin{array}{cc}
\includegraphics[width=0.5\textwidth]{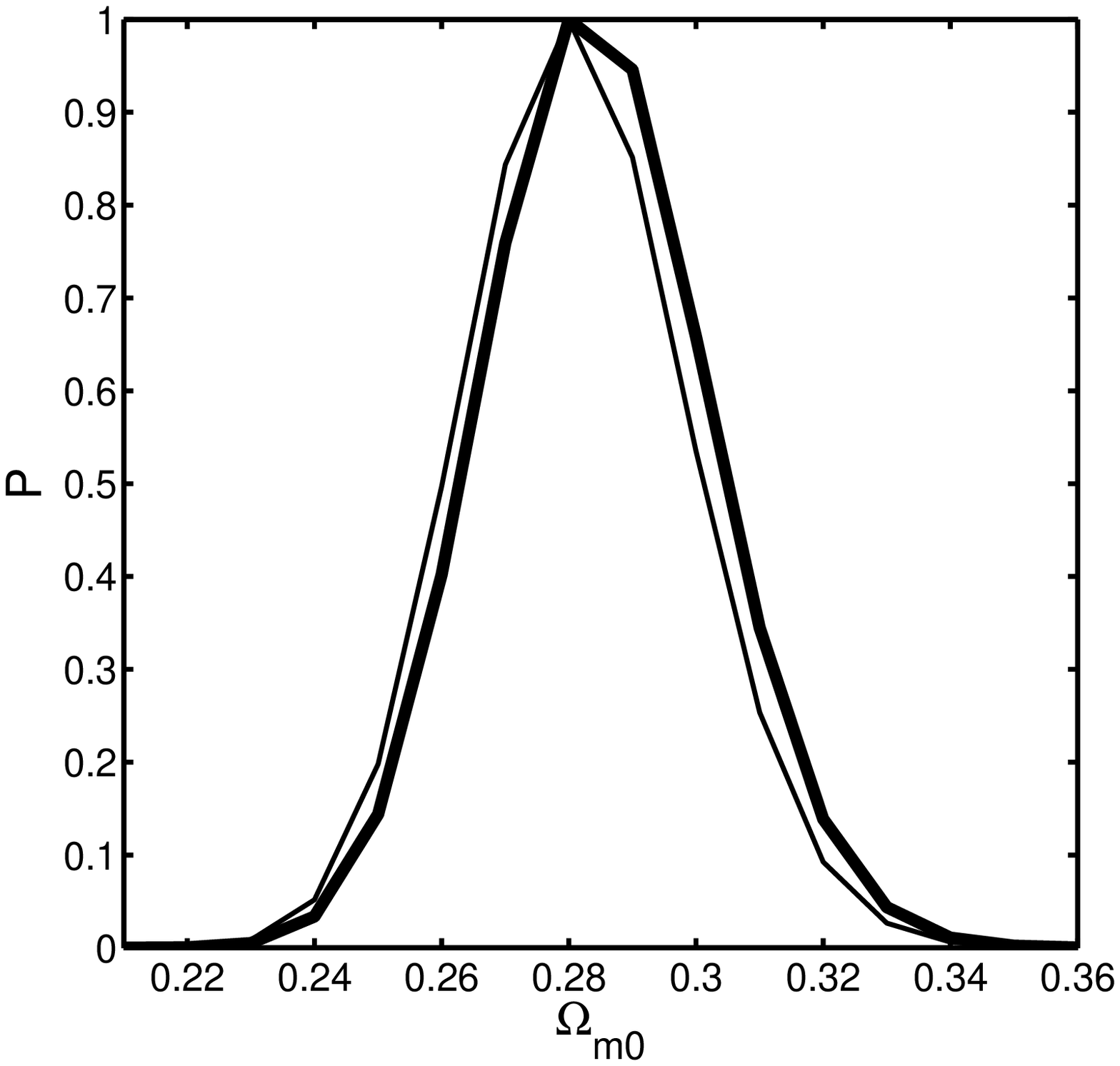}\\
\includegraphics[width=0.5\textwidth]{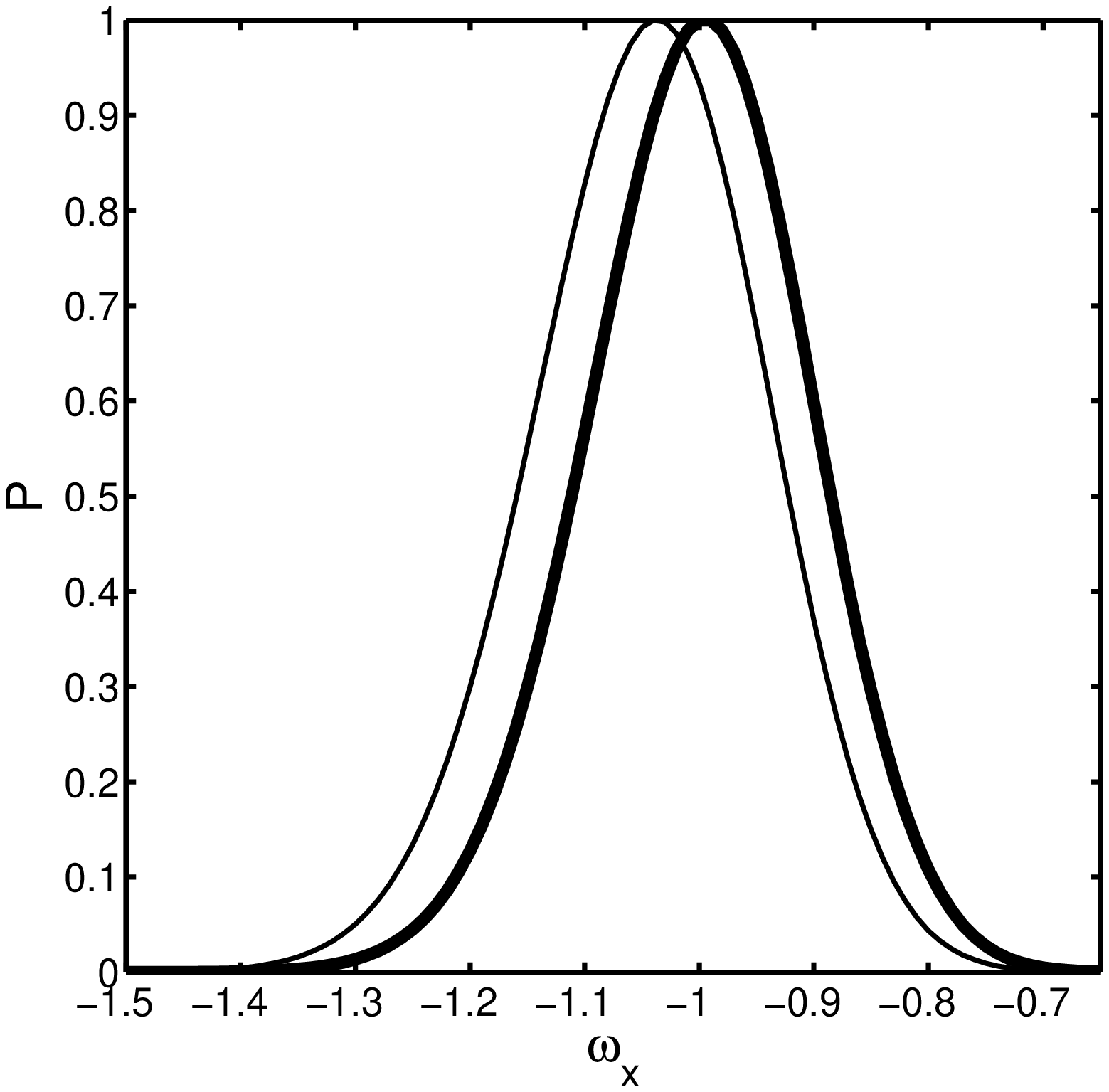}
\end{array}$
 \caption{One-dimensional marginalized distribution probabilities
of the cosmological parameters for the XCDM parametrization. Thick
(thin) lines are results from a joint analysis of the BAO and SNeIa
(with systematic errors) data, with (and without) the ADD data.}
\label{fig:Pro_XCDM}
\end{figure}

\begin{figure}[t]
\centering $\begin{array}{cc}
\includegraphics[width=0.5\textwidth]{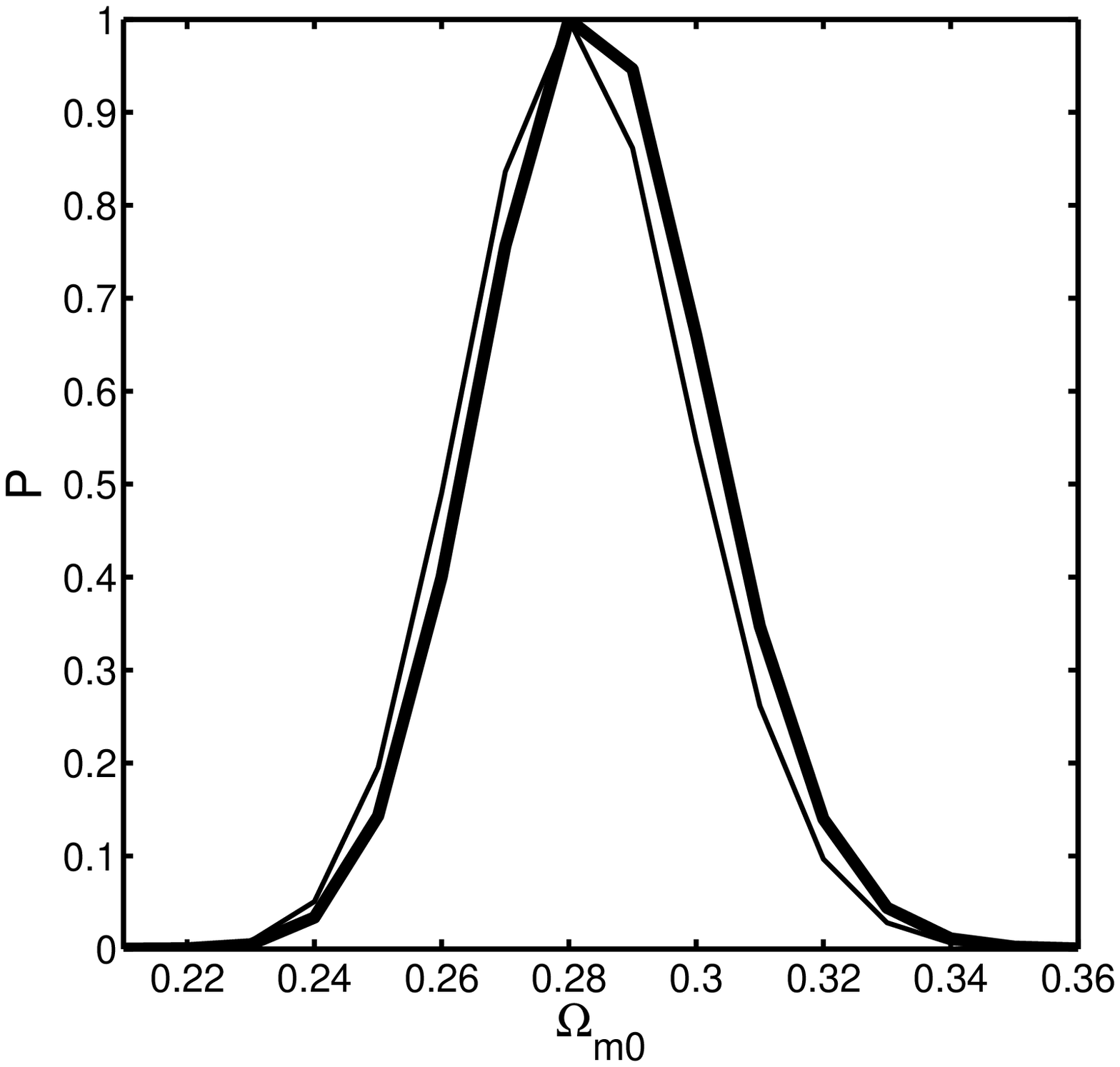}\\
\includegraphics[width=0.5\textwidth]{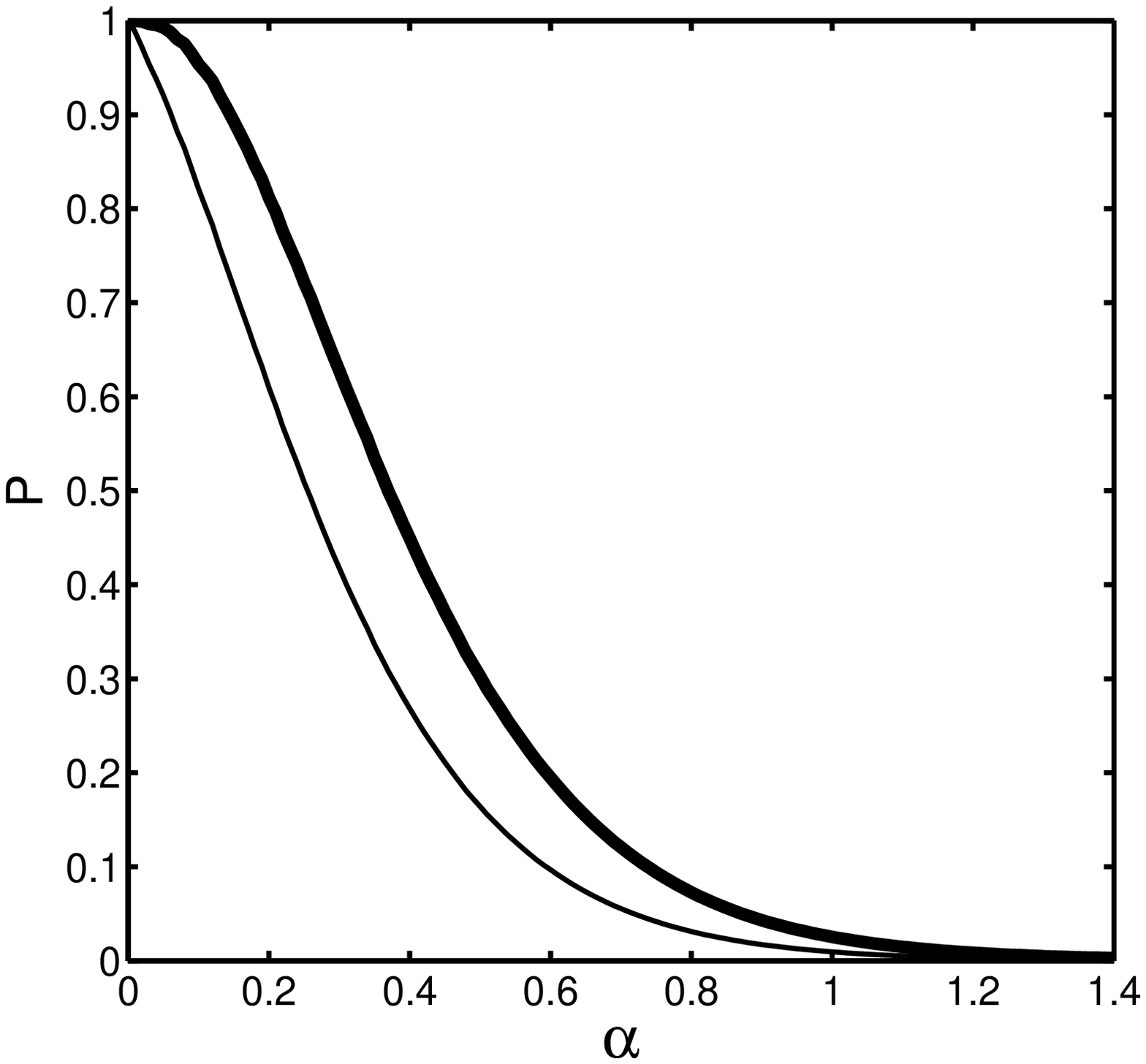}
\end{array}$
 \caption{One-dimensional marginalized distribution probabilities
of the cosmological parameters for the $\phi$CDM  model. Thick
(thin) lines are results from a joint analysis of the BAO and SNeIa
(with systematic errors) data, with (and without) the ADD data.}
\label{fig:Pro_phiCDM}
\end{figure}

Figures \ref{fig:Pro_LCDM}---\ref{fig:Pro_phiCDM} display the
one-dimensional marginalized distribution probabilities of the
cosmological parameters for the three cosmological models considered
in this work, derived from a joint analysis of the BAO and SNeIa
data, as well as from a joint analysis of the BAO, SNeIa and ADD
data. The marginalized 2 $\sigma$ intervals of the cosmological
parameters are presented in Table \ref{tab:intervals}.


\begin{table}
\vspace*{-12pt}
\begin{center}
\begin{tabular}{lcc}
\hline\hline

Model & BAO + SNeIa  & ADD + BAO + SNeIa\\
\hline

$\Lambda$CDM & $0.24<\Omega_{m0}<0.33$
& $0.24<\Omega_{m0}<0.33$  \\
& $0.5<\Omega_{\Lambda}<0.97$ & $0.46<\Omega_{\Lambda}<0.93$\\
\hline

 XCDM & $0.24<\Omega_{m0}<0.33$
& $0.24<\Omega_{m0}<0.33$\\
&$-1.30<\omega_X<-0.80$ & $-1.25<\omega_X<-0.77$\\
\hline

$\phi$CDM & $0.24<\Omega_{m0}<0.33$
& $0.24<\Omega_{m0}<0.33$ \\
& $0<\alpha<0.73$ & $0.01<\alpha<0.89$ \\

\hline
\end{tabular}
\end{center}
\caption{Two standard deviation bounds on cosmological
parameters.}\label{tab:intervals}
\end{table}

The combination of BAO and SNeIa data gives tight constraints on the
cosmological parameters. Adding the currently-available galaxy
cluster ADD data to the mix does shift the constraint contours,
however the effect is not large. Current ADD data do not have enough
weight to significantly affect the combined BAO and SNeIa results.
The ADD data have approximately the same weight as
currently-available gamma-ray burst luminosity measurements
(\cite{Samushia&Ratra2010}, Figs.\ 4---6 and 10---12).

Clearly, the observational data considered here are very consistent
with the predictions of a spatially-flat cosmological model with
energy budget dominated by a time-independent cosmological constant.
However, the data do not rule out time-evolving dark energy,
although they do require that it not vary rapidly.

\section{Conclusion}
\label{summary}

We have shown that the galaxy cluster angular size versus redshift
data of Ref.\ \cite{Bonamente2006} can also be used to constrain
dark energy model
cosmological parameters. The resulting constraints are compatible
with those derived from other current data, thus strengthening
support for the current ``standard'' cosmological model. The ADD
constraints are approximately as restrictive as those that follow
from currently available gamma-ray burst luminosity data, strong
gravitational lensing measurements, or lookback time (or Hubble
parameter) observations. They are, however, much less restrictive
than those that follow from a combined analysis of BAO peak length
scale and SNeIa apparent magnitude data.

The spatially-flat $\Lambda$CDM model, currently dominated by a
constant cosmological constant, provides a good fit to the data we
have studied here. However, these data do not rule out a
time-evolving dark energy.

We anticipate that near-future angular size data will provide
significantly more restrictive constraints on cosmological
parameters. In conjunction with other observations, this angular
size data will prove very useful in pinning down parameter values of
the ``standard'' cosmological model.

\acknowledgments
Yun Chen thanks Rodrigo Holanda, Zong-Hong Zhu and Lado Samushia for
their generous and helpful advice. YC was supported by the China
State Scholarship Fund No.\ 2010604111 and the Ministry of Science
and Technology national basic science program (project 973) under
grant No.\ 2007CB815401. BR was supported by DOE grant
DEFG03-99EP41093.


\begin{thebibliography}{99}

\bibitem{Bonamente2006} M. Bonamente, et al. , \apj 647 (2006) 25-54.

\bibitem{frieman08} J. A. Frieman, in: American Institute of Physics Conference Series,
volume 1057 of American Institute of Physics Conference Series, pp.
87-124.
\bibitem{ratra08} B. Ratra, M. S. Vogeley, \pasp 120 (2008) 235-265.
\bibitem{caldwell09} R. R. Caldwell, M. Kamionkowski, Annual Review of Nuclear and Particle Science 59 (2009) 397-429.
\bibitem{sami09}M. Sami, Curr.~Sci.~ 97 (2009) 887.
\bibitem{Bartelmann2010} M. Bartelmann, Reviews of Modern Physics 82 (2010) 331-382.
\bibitem{cai10} Y.-F. Cai, et al. , Physics Reports 493 (2010) 1-60.
\bibitem{brax09} P. Brax, arXiv:0912.3610
\bibitem{wei11} H. Wei, \plb 695 (2011) 307-311.
\bibitem{jamil10} M. Jamil, E. N. Saridakis, JCAP 7 (2010) 28.
\bibitem{maggiore11} M. Maggiore, Phys. Rev. D 83 (2011) 063514.
\bibitem{dutta10}S. Dutta, R. J. Scherrer, Phys. Rev. D 82 (2010) 043526.
\bibitem{shao10} S.-H. Shao, P. Chen, Phys. Rev. D 82 (2010) 126012.
\bibitem{lepe10}S. Lepe, F. Pe~na, European Physical Journal C 69 (2010) 575-579.
\bibitem{sloth10}M. S. Sloth, International Journal of Modern Physics D 19 (2010) 2259-
2264.
\bibitem{liu10}D.-J. Liu, Phys. Rev. D 82 (2010) 063523.
\bibitem{honorez10}L. Lopez Honorez, et al. , JCAP 9 (2010) 29.
\bibitem{peebles84}P. J. E. Peebles, Astrophys. J 284 (1984) 439-444.
\bibitem{Peebles&Ratra2003} P. J. Peebles, B. Ratra, Reviews of Modern Physics 75 (2003) 559-606.
\bibitem{Perivolaropoulos2010}L. Perivolaropoulos, Journal of Physics Conference Series 222 (2010)
012024.
\bibitem{Ratra&Peebles1988}B. Ratra, P.~J.~E. Peebles, Phys. Rev. D 37 (1988)
3406.
\bibitem{Riess1998}A. G. Riess, et al. , Astron. J 116 (1998) 1009-1038.
\bibitem{ratra91}B. Ratra, Phys. Rev. D 43 (1991) 3802-3812.
\bibitem{Peebles&Ratra1988}P. J. E. Peebles, B. Ratra, Astrophys. J 325 (1988) L17-L20.
\bibitem{Jassal10} H. K. Jassal, J. S. Bagla, T. Padmanabhan, Mon.\ Not.\ R.\
Astron.\ Soc.\ 405 (2010) 2639-2650.
\bibitem{wilson06} K. M. Wilson, G. Chen, B. Ratra, Modern Physics Letters A 21 (2006)
2197-2204.
\bibitem{Davis2007} T. M. Davis, et al. , Astrophys. J 666 (2007) 716-725.
\bibitem{allen08} S. W. Allen, et al. , Mon.\ Not.\ R.\ Astron.\ Soc.\
383 (2008) 879-896.
\bibitem{Perlmutter1999} S. Perlmutter, et al. , Astrophys. J 517 (1999) 565-586.
\bibitem{shafieloo09}
 A. Shafieloo, V. Sahni, A. A. Starobinsky, Phys. Rev. D 80 (2009)
101301.
\bibitem{holsclaw10}T. Holsclaw, et al. , Physical Review Letters 105 (2010) 241302.
\bibitem{ratra99}B. Ratra, et al. , Astrophys. J 517 (1999) 549-564.
\bibitem{podariu01b} S. Podariu, et al. , Astrophys. J 559 (2001) 9-22.
\bibitem{Spergel2003}D. N. Spergel, et al. , ApJS 148 (2003) 175-194.
\bibitem{Komatsu2009} E. Komatsu, et al. , ApJS 180 (2009) 330-376.
\bibitem{Komatsu2011}E. Komatsu, et al. , ApJS 192 (2011) 18-+.
\bibitem{chen03b}G. Chen, B. Ratra, Publ.Astron.Soc.Pac. 115 (2003) 1143-1149.
\bibitem{Percival2007}W. J. Percival, et al. , Mon.\ Not.\ R.\ Astron.\ Soc.\ 381 (2007) 1053-1066.
\bibitem{gaztanaga09}E. Gazta\~{n}aga, A. Cabr\'{e}, L. Hui, Mon.\ Not.\ R.\
Astron.\ Soc.\ 399 (2009) 1663-1680.
\bibitem{samushia09b} L. Samushia, B. Ratra, Astrophys. J 701 (2009) 1373-1380.
\bibitem{wang10a} Y. Wang, Modern Physics Letters A 25 (2010) 3093-3113.
\bibitem{podariu01a} S. Podariu, P. Nugent, B. Ratra, Astrophys. J 553 (2001) 39-46.
\bibitem{samushia11} L. Samushia, et al. , Mon.\ Not.\ R.\ Astron.\ Soc.\ 410 (2011) 1993-2002.
\bibitem{wang10b} Y. Wang, et al. , Mon.\ Not.\ R.\ Astron.\ Soc.\ 409 (2010) 737-749.
\bibitem{Samushia&Ratra2008} L. Samushia, B. Ratra, ApJL 680 (2008) L1-L4.
\bibitem{ettori09}S. Ettori, et al. , Astron.Astrophys. 501 (2009) 61-73.
\bibitem{schaefer07} B. E. Schaefer, Astrophys. J 660 (2007) 16-46.
\bibitem{liang08} N. Liang, S. N. Zhang, in: American Institute of Physics Conference
Series, volume 1065 of American Institute of Physics Conference
Series, pp. 367-372.
\bibitem{wang08} Y. Wang, Phys. Rev. D 78 (2008) 123532.
\bibitem{Samushia&Ratra2010}L. Samushia, B. Ratra, Astrophys. J 714 (2010) 1347-1354.
\bibitem{courtin11}J. Courtin, et al. , Mon.\ Not.\ R.\ Astron.\ Soc.\ 410 (2011) 1911-1931.
\bibitem{baldi10} M. Baldi, Astron. J 411 (2011) 1077-1103.
\bibitem{basilakos10}S. Basilakos, M. Plionis, J. Sol¦Ìa, Phys. Rev. D 82 (2010) 083512.
\bibitem{chae02}K.-H. Chae, et al. , Phys. Rev. Lett. 89 (2002) 151301.
\bibitem{chae04} K.-H. Chae, et al. , ApJL 607 (2004) L71-L74.
\bibitem{lee07} S. Lee, K.-W. Ng, Phys. Rev. D 76 (2007) 043518.
\bibitem{yashar09}M. Yashar, et al. , Phys. Rev. D 79 (2009) 103004.
\bibitem{capozziello04} S. Capozziello, et al. , Phys. Rev. D 70 (2004) 123501-+.
\bibitem{simon05} J. Simon, L. Verde, R. Jimenez, Phys. Rev. D 71 (2005) 123001.
\bibitem{Samushiaetal2010}L. Samushia, et al. , Physics Letters B 693 (2010) 509-514.
\bibitem{dantas11} M. A. Dantas, et al. , Physics Letters B 699 (2011) 239-245.
\bibitem{Samushia&Ratra2006} L. Samushia, B. Ratra, ApJL 650 (2006) L5-L8.
\bibitem{samushia07} L. Samushia, G. Chen, B. Ratra, arXiv:0706.1963.
\bibitem{fernandez08} E. Fernandez-Martinez, L. Verde, JCAP 8 (2008) 23.
\bibitem{yang10}R.-J. Yang, S. N. Zhang, Mon.\ Not.\ R.\ Astron.\ Soc.\ 407 (2010) 1835-1841.
\bibitem{gurvits99} L. I. Gurvits, K. I. Kellermann, S. Frey, Astron. Astrophys. 342 (1999)
378-388.
\bibitem{guerra00} E. J. Guerra, R. A. Daly, L. Wan, Astrophys. J 544 (2000) 659-670.
\bibitem{chen03a} G. Chen, B. Ratra, Astrophys. J 582 (2003) 586-589.
\bibitem{podariu03} S. Podariu, et al. , Astrophys. J 584 (2003) 577-579.
\bibitem{khokhlov2011}D. L. Khokhlov, Int. J Mod. Phys. D 20 (2011) 1167-1169.
\bibitem{debernardis06} F. de Bernardis, E. Giusarma, A. Melchiorri, International Journal of Modern Physics D 15 (2006) 759-766.
\bibitem{holanda08} R. F. L. Holanda, J. V. Cunha, J. A. S. Lima, arXiv:0807.0647.
\bibitem{lima10} J. A. S. Lima, R. F. L. Holanda, J. V. Cunha, in: American Institute of
Physics Conference Series, volume 1241 of American Institute of
Physics Conference Series, pp. 224-229.
\bibitem{li11}Z. Li, P. Wu, H. Yu, ApJL 729 (2011) L14.
\bibitem{liang11}N. Liang, S. Cao, Z.-H. Zhu, arXiv:1104.2497.
\bibitem{meng11} X.-L. Meng, T.-J. Zhang, H. Zhan, arXiv:1104.2833.
\bibitem{chen03} G. Chen, J. R. Gott, III, B. Ratra, Publ. Astron. Soc. Pac. 115 (2003) 1269-1279.
\bibitem{Percival2010} W. J. Percival, et al. , Mon.\ Not.\ R.\ Astron.\ Soc.\
401 (2010) 2148-2168.
\bibitem{samushia09a} L. Samushia, B. Ratra, Astrophys. J 703 (2009) 1904-1910.
\bibitem{Eisenstein2005}D. J. Eisenstein, et al. , \apj 633 (2005) 560-574.
\bibitem{Amanullah2010}R. Amanullah, et al. , \apj 716 (2010) 712-738.
\bibitem{DiPietro&Claeskens2003}E. di Pietro, J.-F. Claeskens, Mon.\ Not.\ Roy.\ Astron.\ Soc.\ 341 (2003) 1299-1310.
\bibitem{Perivolaropoulos2005}L. Perivolaropoulos, Phys. Rev. D 71 (2005) 063503.
\bibitem{Nesseris&Perivolaropoulos2005} S. Nesseris, L. Perivolaropoulos, Phys. Rev. D 72 (2005) 123519.
\bibitem{Kowalski2008}M. Kowalski, et al. , Astrophys. J 686 (2008) 749-778.
\end{thebibliography}
\end{document}